# Review of the Plan for Integrating Big Data Analytics Program for the Electronic Marketing System and Customer Relationship Management: A Case Study XYZ Institution

**Idha Sudianto**

Magister of Management Post-Graduate Study, STIE Perbanas Surabaya, Indonesia

**Abstract**
This research aims to explore business processes and what the factors have major influence on electronic marketing and CRM systems? Which data needs to be analyzed and integrated in the system, and how to do that? How effective of integration the electronic marketing and CRM with big data enabled to support Marketing and Customer Relation operations. Research based on case studies at XYZ Organization: International Language Education Service in Surabaya. Research is studying secondary data which is supported by qualitative research methods. Using purposive sampling technique with observation and interviewing several respondents who need the system integration. The documentation of interview is coded to keep confidentiality of the informant. Method of extending participation, triangulation of data sources, discussions and the adequacy of the theory are uses to validate data. Miles and Huberman models is uses to do analysis the data interview. Results of the research are expected to become a holistic approach to fully integrate the Big Data Analytics program with electronic marketing and CRM systems.

## I. INTRODUCTION

Emyana Ruth ErithaSirait (2016: 114), data has an important role in strategic decision making, unfortunately the application of Big Data Analytics is not popular in Indonesia. So far, the three main user of Big Data in Indonesia are telecommunications companies, banks, and low-cost consumer goods companies, such as foods and beverages (consumer goods). For example (SAP Indonesia Community' data): XL Axiata, Telkom, Indosat, Unilever, Sampoerna, BCA has integrated Big Data Analytics with Electronic Marketing and CRM using SAP ERP software, the results are very satisfying, on the other hand it requires very large financial license. All these companies have unlimited resources, very difficult for small or medium-sized companies, and formal and non-formal educational institutions to implement it. Although, there are various types open source licensed and free of Big Data applications, Electronic Marketing, and CRM.

Free licensed Big Data Analytics program does not make it easy for companies and agencies easily to adopt it. Some of these applications include Apache Hadoop, Oddoo, and Tableau. The limited research, article journals, and references in Indonesia and the limitations of HR skills are the main issues of implementing Big Data. Another obstacle is that if someone successfully integrates the big data analysis program, its nature is still general, not specific to the user's needs. In other words, it can be concluded that software is not able to adjust to the users need, but users must adjust to the processes contained in the analysis software. Inductive thinking process is needed which is aligned with further academic studies, then to do research with a qualitative method approach. Research studies are needed to answer the question: Does the company really need integration to increase growth? What variables must be considered in the integration process? How the effectiveness of system integration can support the operation of electronic marketing and CRM.

The case study research was conducted at the XYZ organization: International Language Education Service. The research objects though have different business lines and segments, but have the same problem, stagnant growth.

Weija Li (2016): The results of digital data analysis can improve marketing strategies, find potential market and can correcting the results of thinking and human analysis (marketers). It is important to combine data analysis with an in-depth understanding of the market in interpreting customer journey. This is in line with the arguments of Edelman and Singer (2015): analyzing and interpreting data across different stages can add more value to consumers, improve brand image and competitive advantage. Georgia Fotaki et all (2013): Business objectivity variables (online marketing and online customer engagement): increase new customers, increase customer satisfaction and loyalty, reduce churn rates, increase sales. Segmentation that supports the company's business processes behavior, demography, loyalty, reference system (positive WOM). Whereas according to Christopher A. Beloin (2018): participation or involvement of students, and customers play an important role in the success of a CRM system. By designing a system that can involve customers such as communication feedback, suggestions, complaints including the customers tracing history will increase customer retention. The CRM system proved to be very effective in supporting the duties of the Customer Relation division in daily operations

## II. THEORETICALFRAMEWORK

Companies who can process and utilize data distribution, will get many benefits for the process of developing and sale their products or services, have competency in maintaining good relationships and customer loyalty. According to the International Data Corporation (2017: 15) report, global electronic data growth has doubled in quantity every two years. It is a strategic challenge to use these to be more useful. Unstructured data compositions are more created or dominated than structured data. Emyana Ruth ErithaSirait (2016: 118) structured data has data types, formats, and structures that have been defined. Data can be in the form of transactional data, OLAP data, traditional RDBMS, CSV files, simple spread sheets. While unstructured data is textual data with an erratic format or no inherent structure, so to make it structured data requires more effort, tools, and time. This data is generated by internet applications, such as log URL data, social media, e-mail, blogs, video, audio and semantic. Since the introduction of Big Data in data collection and analysis, this technique has begun to be compared with conventional methods previously carried out, such as surveys.

### 2.1 Big Data

McKinsey Global (2011), Big Data can be defined by data that has scale (volume), distribution (velocity), variety that is very large, and or enduring, thus requiring the use of technical architecture and innovative analytical methods to gain insight who can provide new business value (meaningful information). And in its development, there are those who call 7V including: Volume, Velocity, Variety, Variability, Veracity, Value, and Visualization, or 10V even more than that. Big data is a term for a large or complex set of data which cannot be handled with conventional computer technology systems (Hurwitz: 2013). Whereas according to Emyana Ruth ErithaSirait (2016: 114) data has an important role in strategic decision making. Therefore, parties who can process and utilize available data in large volumes, vary in diversity, high complexity and high speed of data additions, can take large profits.

### 2.2 Electronic Marketing

Judy Strauss and Raymond Frost (2014: 23), electronic marketing or e-marketing is the use of information technology for marketing activities, as well as the process of creating, communicating, sending, and exchanging offers that have value for customers, clients, partners and the wider community. Simplified e-Marketing is the result of information technology that is adopted for traditional marketing methods. Whereas according to Candra Ahmadi and Dadang Hermawan (2013: 186), e-Marketing is part of e-Business that uses electronic media to conduct marketing activities to achieve marketing goals. There are various forms of e-Marketing such as internet marketing, interactive marketing, and mobile marketing. Based on the terminology above, it can be concluded that electronic marketing (e-Marketing) or electronic marketing is also known as Internet Marketing, Web Marketing, Digital Marketing or Online Marketing.

The marketing mixes on electronic marketing according to Kalyanam and McIntyre (2002), there are 11 electronic marketing functions that are described in the form of a marketing mix element. The configuration of valuation, facilitation and symbolization is the basic function of electronic marketing that is mapped into products, prices, places and promotions as separate electronic marketing elements. Nine out of 11 are classified as basic functions, while 7 functions moderate other effects (overlapping). The non-overlapping function is placed on the surface of the cube. Functions that moderate other functions are placed at the bottom of the cube to illustrate that most of these functions operate by moderating functions on the surface to add moderation to each other. The results of the electronic marketing mix can be formulated: 4Ps + P2C2S3.

> 4S = Product, Price, Promotion, Place.
> P2 = Personalization, Privacy.
> C2 = Customer Service, Community.
> S3 = Site, Security, Sales Promotion.

Kalyanam and McIntyre also identified electronic marketing tools that were grouped according to the electronic marketing mix. By mapping the tools in each of the electronic marketing mix functions, will be able to assist the implementation or implementation of the electronic marketing system itself. There are four domains of online marketing according to Philip Kotler and Gary Armstrong (2014: 526): Business to Consumer (B2C), Business to Business (B2B), Consumer to Consumer, and Consumer to Business (C2B). Business to Consumer (B2C) can be interpreted as a process for companies selling products and services online directly to end user. Business to Business online marketing (B2B), marketers use websites, electronic mail, social media, and other online resources to find new corporate customer business opportunities, sell to existing customers, and serve customers effectively and efficiently. Customer to Customer (C2C), with internet media between customers can be interconnected to exchange offers. The fourth domain is Consumer to Business (C2B), consumers can directly go to the company's website page to find goods or services that suit their needs. By understanding the domain of online marketing, companies and marketers can easily develop strategic planning for electronic marketing systems.

Customer involvement is needed by companies in designing and creating promotional content. Participation occurs when internet users connect or collaborate with brands, companies. This can create an emotional connection with customers, because their needs feel fulfilled and cared for. By accommodating customer needs, characteristics, and behavior, it is certain that every marketing content and product offer will be well received by the customer. According to Ari Setyaningrum et al (2015: 390 - 396), the stages in conducting internet marketing/ online marketing are:

1. Making a website.
2. Online advertising or promotion.
3. Making social networks or online communities.
4. Use of e-mail.
5. Mobile marketing: based on SMS blast, promotional content that can be accessed by smartphones.

## 2.3 Customer Relationship Management

Francis Buttle (2015: 16), Customer Relationship Management or CRM is the core of business strategies that are integrated with internal processes and functions, and outside networks. To create and convey value to customers. This underlies the high quality of customer relations and can be achieved by implementing information technology systems. There are three types of CRM:

1. Strategic CRM: focused on developing products / services in accordance with the company's business culture that is oriented to the needs and desires of customers, to maintain customer loyalty by creating and delivering more value compared to competitors.
2. Operational CRM: focuses on business automation processes desired by customers. Through a marketing automation system, sales force automation, and service automation.
3. Analytical CRM: focused on the process of obtaining, storing, processing, integrating, distributing and using customer data, used to increase value from both the customer and the company.

There are four comprehensive models of CRM that have been developed to date: IDIC Models, Value Chain CRM, Payne and Frows 5-process models, and Gartner competency models. IDIC model is the most suitable for thesis research. Here are four processes for building a CRM based on one-to-one relationships with customers:

1. Identification: who is the company's customer and how to understand them well.
2. Differentiation: identify which customers have great potential and have more prospects in the future.
3. Interaction with customers make sure the company can understand customer desires, customer relationships with other suppliers and brands.
4. Customize the types of offers and means of communication to ensure that customer desires can be fulfilled.

By mapping interactions between customers and companies, a clear business process and interaction will be obtained to carry out an appropriate CRM system planning analysis. Analytical CRM by utilizing good customer data structured in internal data centers and unstructured data, to be processed and analyzed will be able to produce a system of relationships between customers and companies that are more custom and specific.

### III. RESEARCH FRAMEWORK

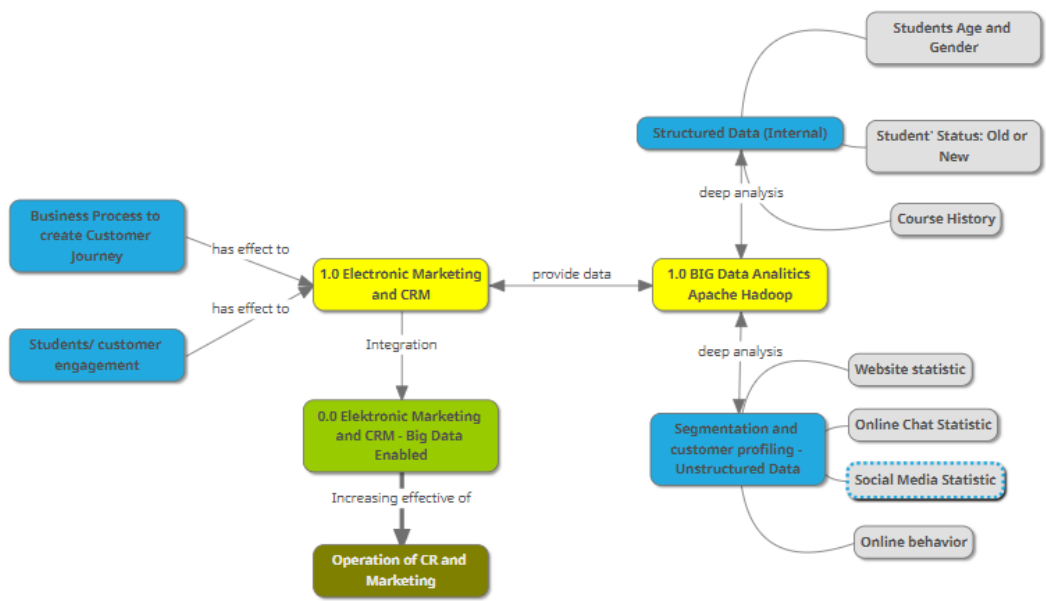

Figure1: Research Framework

The following are some of the main points of thesis research based on theoretical Big Data studies, Electronic Marketing and Customer Relationship Management Systems:

1. Big Data Analytics

   A. Tools and methods: Apache Hadoop.

   B. Data processed: structured internal data and unstructured data (Facebook, Instagram, YouTube Channel, Chat Analytics, and web analytics). Analysis of electronic marketing reports, Analytical CRM: email, SMS and online chat.

   C. Output of the analysis process: trend data and analysis of the effectiveness of marketing and CRM systems that have been carried out by the company.

2. Electronic Marketing

   A. Marketing tools and methods: websites, online promotions through social networks, e-mail marketing, SMS blasts.
   B. Data processed: ---
   C. Output: statistical reports of customers who receive promotions from the system.

3. Customer Relationship Management

A. Method: Analytical CRM, Operational CRM
B. Data processed: structured internal data and unstructured data (Facebook, Instagram, YouTube Channel, Chat Analytics, and web analytics).
C. Output: a custom and personal relationship system through electronic mail, SMS and Line for Business and WhatsApp for Business media.

## IV. RESEARCH DESIGN

This research uses qualitative research methods. In each system development plan always requires descriptive data in the form of structured interviews from each party in the research subject, to obtain factual data in more depth (natural setting). Direct observation, observation, checking or discussion is also needed to further enrich the research data on a problem. According to I Wayan Suwendra (2018: 8), one of the benefits of qualitative research is the improvement of practice, case study research will trace the planning, processes and follow-up of a program so that it is very valuable in improving a practice. The background of the problem or factual problem is the starting point of this research.

Table 1: List of Informants

| No | Criteria | Informan | Position |
|---|---|---|---|
| 1 | Fully responsibile for company development and the growth of organization. | Mr. FOB | Branch Manager |
| 2 | Extensive academic and professional insight into the organization's marketing system and have skills in digital marketing planning. | Ms. SN | Head of Marketing and Digital Content |
| 3 | Have competence to maintaining good relationships with students and parents. | Mr. HDW | Head of Customer Relation Div |
| 4 | PIC customer *relationship* and marketing. | Ms. RD dan Mr. SIA | Customer Relation Staff |
| 5 | Having knowledge of Big Data, IT, Electronic system development (websites, Line Apps, social media). | Mr. DEP | Technical and Information System Officer |
| 6 | Have global insight regarding development of the English language education system. | Mr. AN | Director of Study |
| 7 | Global insight for the Business English Program development, directly interacting with clients. | Mr. FJ | Assistant Director of Study |

Problem identification is carried out in weekly meetings. Given the data confidentiality factor, the meeting notes cannot be fully described in this study. To maintain the validity of the data and research issues, here are some reflective notes that have been carried out to reduce the problem, so that research remains focused on its objectives:

1. Initial (informal) interview with Ms. SN as the head of the Digital Marketing Promotion division: "*We need an integrated system that is able to analyze the effectiveness of digital marketing, knowing the trends in the informal education market in accordance with organizational segmentation. From the marketing and CRM systems, how many students register*?".

2. Initial (informal) interview with Mr. RA as a Graphic Designer and Digital Content Designer, is also a staff of Ms. SN: "*In making video or digital graphic content, we should look at the customer side, what they need is not just designing a content that emphasizes organizational services or products. Likewise, it should involve students and parents in creating content such as: testimonials, interactive videos* ".

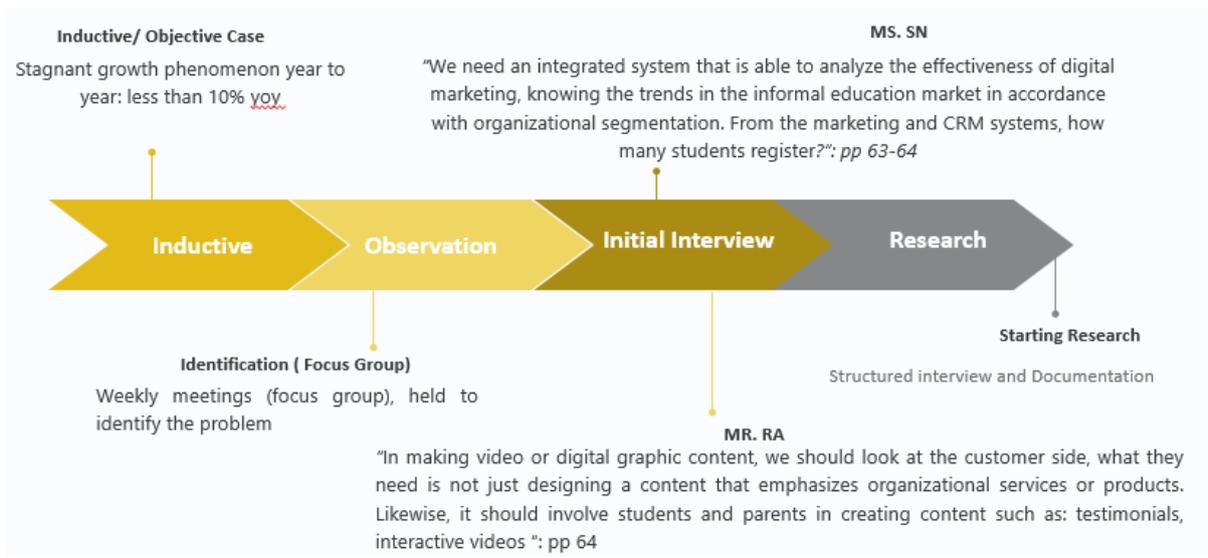

Figure 2: Research Stages

Informal interviews are conducted, followed by concluding the results of the weekly meeting, especially on points relating to this research study: Marketing, Customer Relations and Program Development. Following is a list of questions contained in the structured interview guide, interviews were conducted at each informant.

**Shows data on growth and number of students over the past five years.**

Q1.1: Do you think our student growth is stagnant, can it be overcome by several policies that have been taken, such as:

    1. Marketing through electronic media, internet and social media: Instagram, Facebook, Twitter and Youtube?

    2. Our CRM system with WA and Line for Business call center methods, email and SMS blasts?

Q1.2: If not, are there business processes or other factors that should be analyzed, so that we can support our digital marketing system and CRM that has been implemented to be more optimal?

Shows web statistics and social media log systems, call centers and e-mail.

Q2.1: The data is a statistical report of our system. Are there other data or variables that need to be included in our market segmentation analysis process?

Q2.2: What do you think by integrating the entire marketing system through technology will simplify the marketing process, can optimize communication facilities to maintain good relations with students and parents / customers?

Q3.1: Already optimizing our current data analysis process? whether in the form of AFAS data, exhibition, event and other electronic inquiry?

Q3.2: To follow up from the analysis of the data, have we done it optimally?

Shows pictures of Research Thinking Framework and Proposition.

Q4: What do you know about Big Data and what is your response regarding the system integration plan by the Technical and Information System Division?

Q5.1: Do you think that our draft system integration system will help Marketing and Customer Relation in the daily work and operational tasks?

Q5.2: From our discussion process are there other process variables or businesses that need to be integrated?

Q5.3: From our study of system integration planning, were there other comments and opinions according to brother's expertise that should have been added or reduced? For the system to be created more optimal.

Q6: Shows the results of the analysis of discussions with informants to other informants who appeared in the previous interview (source informant anonymized / hidden) --- the process of checking peers and snowballing, Triangulation Process (Miles Huberman) begins in the second interview and so on.

Sugiyono (2018: 222-225), there are two main things that affect the quality of research results, namely, the quality of the instrument and the quality of data collection. In qualitative research, data collection is carried out in natural settings, primary data sources, and more data collection techniques in participant observation, in-depth interviews, and documentation. The following are some of the data collection methods that will be carried out in this study:

1. Complete participatory observation
   The researcher is involved with the daily activities of the subjects being studied or used as research data sources. In collecting data, researchers have been fully involved in what is done by the data source, the atmosphere is natural, and researchers are not seen doing research. With the activities of researchers as employees and always having good relations with the subject, observations can be made directly to the stages of selected observation (observation mini tour). Researchers can describe the focus found so that the data is more detailed.
2. Structured interviews
   Researchers use this interview technique as a method of data collection, because they already know for certain about the information to be obtained. Respondents were given the same question, the researcher recorded and documented. Research instruments such as corporate secondary data, images and digital voice recorder are used to help the interview process run smoothly.
3. Documentation
   Data in the form of photo documentation, and electronic files recorded, then partners will be made into a complete transcript with coding to maintain the confidentiality of research subject data.

## V.     DATA ANALYSIS

In qualitative research, the main criteria for research data are valid, reliable and objective. Test the validity of the data in this research uses a credibility test which includes:

1. Extension of observation or participation
   Focused on testing the data obtained either directly to the main informant or additional informants according to the snowball method. A certificate will be made as the validity of the test extension to be attached to the attachment to the research study.

2. Triangulation

   Using the source triangulation method, by checking the correctness of the data obtained through various sources that are relevant to the subjects in the same division. Focus on digital marketing, CRM and expected benefits with Big Data integration.

3. Checking or discussing with colleagues.

4. Using enough references.
   Using supporting tools in the form of recording interviews with digital voice tapes, photos, documentation of notes made by researchers directly at the interview (verified subjects by signing or initialing documentation). Sugiyono (2018: 245), analysis of qualitative data is inductive in nature, namely an analysis based on the data obtained, then developed into a hypothesis. Based on the hypotheses formulated based on the data, then the data is searched again repeatedly so that it can then be concluded whether the hypothesis is accepted or rejected based on additional data collected. When based on data that can be collected repeatedly with triangulation techniques, it turns out the hypothesis is accepted, then the hypothesis develops into a theory. The theory was used for the next stage, namely the process of integrating Big Data with marketing systems and corporate CRM. So that system integration is in accordance with the subject's objective needs and is expected to improve or answer existing problems.

In this study, data analysis was used in the field with the models of Miles and Huberman. Data analysis is carried out at the time of data collection, and after completion in a certain period. At the interview, the researcher conducted an analysis of the answers of the informants. If the answers interviewed after being analyzed feel unsatisfactory, the researcher will continue the question again, to a certain extent, until data is deemed credible. The activity in data analysis follows the flow model that is presented by Miles and Huberman.

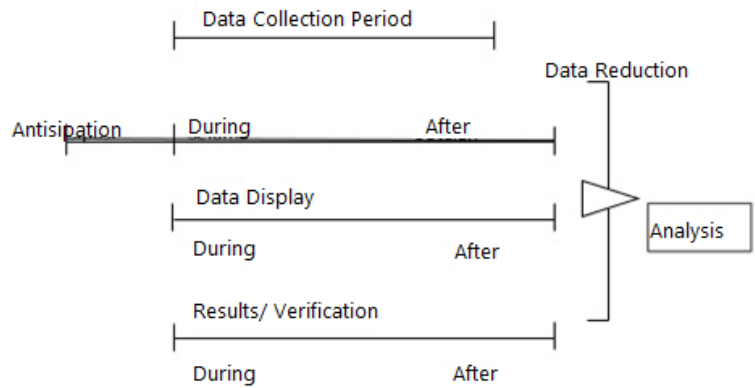

Figure 3: Data Analytics Component (*flow model*)

Data analysis activities are divided into three: data reduction, data display, and conclusion drawing / verification. With the logic scheme in Figure 3.3:

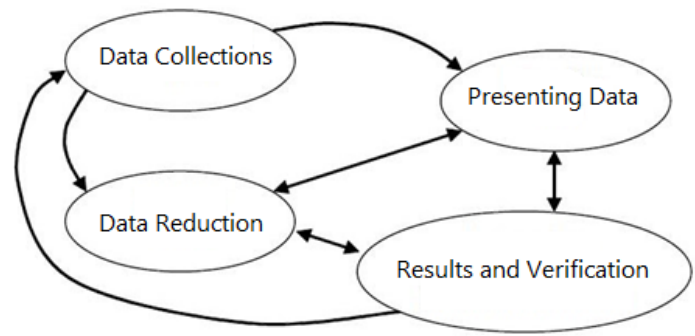

Figure 4: Data Analytics Component (*interactive model*)

With Data Reduction, there are quite a lot of data obtained from the field, so that it needs to be carefully and in detail. Reducing data means summarizing, choosing the main things, focusing on important things, to look for themes and data patterns. After the data is reduced, the next step is to display data. Flowchart began, and narrative texts are the output of this process. Conclusion data in the conclusion drawing / verification will be displayed with images or charts supported by credible data will be the conclusion in this research study.

## VI. PREPOSITION

In accordance with the results of pre-research interviews, inductive thinking supported by academic studies and journal articles, the following are research propositions: The business process of electronic marketing through social media with content that involves the participation of students will be able to create customer journey. Factors in feedback communication, suggestions, complaints, including the tracing of customer history (parents and students) are very influential in electronic marketing and CRM. Need to analyze secondary sequential data in the form of age and sex of students, student status (new or old), and student course history. Big Data Apache Hadoop is needed to analyze website data, online chat, social media, and online behavior from students. Sequential and big data analysis is then comparative to get customer profiling and customization relationships. By creating interactive electronic marketing content, communication feedback, availability of customer profiling and customization relationships. Can improve the operational effectiveness of the Customer Relation and Marketing department.

## VII. RESULTS

The extending participation, triangulation of data sources, discussions and Miles and Huberman models are used to do analysis the data interview. Here the results of this research study:

1. Quality of Big Data Integration Services
   Easy of data access, responsive integration of the Electronic Marketing and CRM systems, and ease of obtaining data are the key point. Besides that, the speed factor for data access and security is important, also.
2. Electronic Marketing and CRM Features.
   The creation of several channels of communication with customers, marketing content match to trends, evaluation and follow-up of communication channels, and the presence of content update features were found as research study in the Electronic Marketing and CRM Features group.
3. Analysis of Other Data (including factors of Customer Engagement and Customer Journey Creation).
   Customer satisfaction and complaint analysis, analysis of potential customer data, market segment evaluation and trend analysis are findings that can answer the question "*What other things need to be analyzed in the integration of this system*?".

## VIII. CONCLUSION

To integrating Big Data Analytics Program for The Electronic Marketing System and Customer Relationship Management, the subject must concern to students and customer engagement for create advertising content met to current trend. Also, the business processes to create Customer Journey has significantly affect to the Electronic Marketing and CRM System runs effectively. Easy to get data and access, security, creating several channel communications trough social media and online chat are another important point were found from this research study. All those points were mentioned by respondents while interview, found in the data reduction from processes of the extending participation.All respondents agreed that Electronic Marketing System and CRM with big data enabled can increasing effectivity operational of Marketing and CRM department.

This research study has been conducting in the International Language Education Service in Surabaya, with a unix customer segmentation and demography. The result is fully supporting the research preposition that proposed in this study. The qualitative research method needs to be conducted to find out how the effectivity of system at all.